\documentclass[a4paper]{spie}
\usepackage{url}
\usepackage{epsfig}
\usepackage{subfigure}
\usepackage{algorithm}
\usepackage{algorithmic}

\renewcommand{\vec}[1]{\underline{\mathbf {#1}}}
\renewcommand{\delta}{z}

\title{Wide spread spectrum watermarking with side information and interference cancellation} 

\author
{
	Ga\"etan Le Guelvouit and St\'ephane Pateux
	\skiplinehalf
	IRISA/INRIA, Campus de Beaulieu, 35042 Rennes Cedex,
	FRANCE
}

\authorinfo{{\tt \{gleguelv, spateux\}@irisa.fr}; phone: +33 2 99 84 25 88; fax: +33 2 99 84 71 71}

\begin{document} 
  	\maketitle 

	\begin{abstract}
		Nowadays, a popular method used for additive
		watermarking is wide spread spectrum. It consists in
		adding a spread signal into the host document. This
		signal is obtained by the sum of a set of carrier
		vectors, which are modulated by the bits to be
		embedded. To extract these embedded bits, weighted
		correlations between the watermarked document and the
		carriers are computed. Unfortunately, even without any
		attack, the obtained set of bits can be corrupted due
		to the interference with the host signal (host
		interference) and also due to the interference with
		the others carriers (inter-symbols interference (ISI)
		due to the non-orthogonality of the carriers). Some
		recent watermarking algorithms deal with host
		interference using side informed methods, but
		inter-symbols interference problem is still open. In
		this paper, we deal with interference cancellation
		methods, and we propose to consider ISI as side
		information and to integrate it into the host signal.
		This leads to a great improvement of extraction
		performance in term of signal-to-noise ratio and/or
		watermark robustness. 
	\end{abstract}

\keywords{Robust watermarking, spread spectrum, side information, interference cancellation}

\section{INTRODUCTION}\label{sect:intro}

	First studies in robust watermarking were mostly empirical.
	The domain became more academic when the watermarking problem
	was considered as communication over a noisy channel~: the
	watermark is a signal to be transmitted through a channel corrupted
	by noise due to the cover signal and attacks. Watermarking
	was then considered as a kind of channel coding. The latest
	contributions then focused on theoretical studies, inspired by
	information theory, but not usable as such. 

	Due to constraints on the embedding distortion (MSE or weighted MSE),
	the power of the transmitted signal is limited.
	The communication channel is noisy due to attacks. It has
	often been modeled as the addition of white Gaussian noise
	(AWGN channel)\cite{Servetto98,Moulin99}. 
	The host signal has then often been considered as a noise
	that limits the performance of the watermarking scheme. But
	recently, it has been shown that watermarking can be regarded as a
	problem of communication with side information~\cite{Cox99}: a
	part of the added noise ({\it i.e.} the host signal) is perfectly
	known during the embedding process. Costa~\cite{Costa83}
	studied this kind of channel and gave a limit of capacity,
	independant of of the host signal. He also exhibited a theoretical algorithm (the
	Ideal Costa Scheme) to reach this limit, considering i.i.d.
	Gaussian signals and AWGN transmission. However since this
	scheme relies on exhaustive search among codevectors,
	practical implementation of this scheme is not realistic. Some
	implementations inspired by the ICS were then proposed, using
	structured codebooks: Eggers's SCS~\cite{Eggers00b} or syndrome
	based codes~\cite{Chou00}.

	Costa's scheme assumes i.i.d. Gaussian signals. Unfortunately,
	real multimedia signals are not so simple. Moreover, attacks may
	be not modeled as simple AWGN channels. Several studies
	proposed to considered  non i.i.d. SAWGN\footnote{Scaling and
	Additive White Gaussian Noise.}
	channels\cite{SuEggersGirod2001,MoulinIvanovic2001,EggBaumlGirod2002}.
	Indeed this class of attacks allows to take into account for
	filtering (such as Wiener filtering for noise removal),
	scaling, addition of noise correlated to the host signal,
	noise from compression\ldots Furthermore, it has been
	shown~\cite{Moulin01,Cohen02} that optimal attacks
	are of the kind SAWGN. In order to use ICS properties,
	watermarking in a linear subspace using wide spread spectrum
	(WSS) has been considered. While our previous
	work~\cite{Guelvouit02} assumes non i.i.d. Gaussian signals,
	thanks to the use of spread transform subspace, 
	projected host signal and attack noise are i.i.d and Gaussian.
	Furthermore, this scheme leads to a practical implementation
	with performances close to optimal~\cite{Pateux03}. However
	adding a watermark in this subspace introduces symbol
	interference due to the non-orthogonality of the carriers used
	for the spread transform. This ISI, like the host signal in
	non-informed watermarking, limits the performance of the
	scheme. 

	This paper deals with a practical and complete informed
	watermarking scheme, using spread spectrum and structured
	codebooks. It also provides a solution to symbol interference
	cancellation. In Sec.~\ref{sec:state}, we recall the
	subspace-based approach and introduce a structured codebook
	based on punctured convolutional codes. In Sec.~\ref{sec:ic},
	we first study two ISI cancellation methods, and we then provide an
	iterative algorithm to consider symbol interference as side
	information, illustrated by experimental results in
	Sec.~\ref{sec:results}. We finally conclude this paper in
	Sec.~\ref{sec:conclusion}. 

\section{SPREAD SPECTRUM FOR SIDE INFORMED WATERMARKING}\label{sec:state}

	We have shown in our previous work~\cite{Guelvouit02,Pateux03}
	a practical scheme that achieves performances close to the
	optimal bounds~\cite{Moulin01}. The watermark is embedded in a linear
	subspace: i.i.d. Gaussian signals are then obtained and ICS
	can be applied. We first recall in this section the original
	Costa's approach. In order to render realistic ICS, we then
	introduce a structured codebook (dirty paper codes) based on
	convolutional codes. We finally describe our WSS-based
	embedding method, optimized using game theory (min-max
	optimization). 

	\begin{figure}
  		\centerline{\epsfig{figure=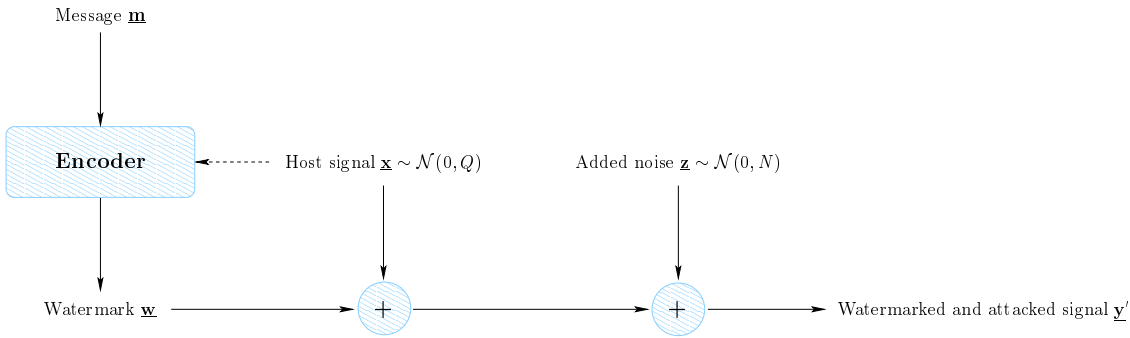,width=15cm}}
	    	\caption{The watermarking channel seen as communication with side information.} 
    		\label{fig:channel}
	\end{figure}

	\subsection{Channel with side information: Costa's approach\label{sec:costa}}

	As seen in the introduction, the watermarking problem can be
	seen as a communication process with side information
	available at the encoder~\cite{Cox99}. This kind of channel
	have been studied by Costa~\cite{Costa83}, which leaded to
	an upper bound of capacity for this kind of channel. 

	Let us consider a $n$-long i.i.d. Gaussian host signal
	$\vec{x}$, whose samples are modeled by $X \sim {\mathcal N}(0, Q)$. This signal
	is perfectly known during the embedding process. We transmit
	our data with a watermark signal $\vec{w} = \left\{ w_1, w_2,
	\ldots , w_n \right\}$ as seen on Fig.~\ref{fig:channel}. The
	energy of $\vec{w}$ is bounded so that 	
	\begin{equation}
		\frac{1}{n} \sum_{i=1}^n w_i^2 \leq P \textrm{.}
		\label{eq:bound}
	\end{equation}
	The transmitted signal is then $\vec{y} = \vec{x} + \vec{w}$.
	This signal is corrupted during the transmission by an added
	Gaussian noise $\vec{\delta}$, modeled by $Z \sim {\mathcal
	N}(0, N)$. Receiver then gets the signal $\vec{y}' =
	\vec{y} + \vec{\delta}$. If we consider this channel as a
	classical Gaussian one, two noises are added to the
	transmitted signal, so the capacity is given by
	\begin{equation}
		C = \frac{1}{2} \log_2 \left[
			1+ \frac{P}{Q+N}
		\right] \textrm{.} \label{eq:classic}
	\end{equation}
	Side information $\vec{x}$ impacts on the performance of the
	system, lowering the capacity. Costa showed that the side
	information does not influence the optimal capacity of the
	channel, {\it i.e.} 
	\begin{equation}
		C = \frac{1}{2} \log_2 \left[
			1+ \frac{P}{N}
		\right] \textrm{.}
		\label{eq:costa}
	\end{equation} 

\begin{figure}
  		\centerline{\epsfig{figure=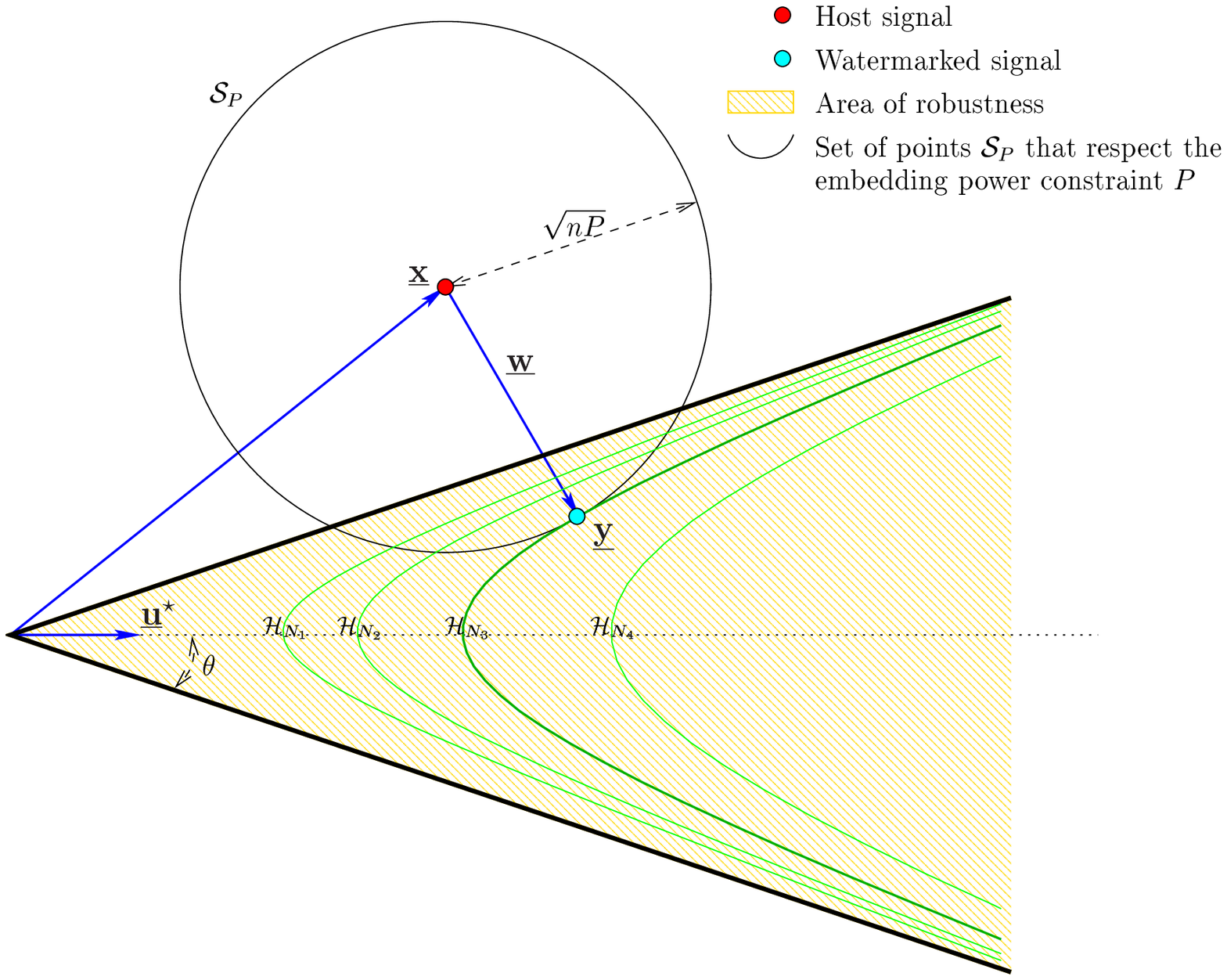,width=9cm}}
	    	\caption{Perturbation of signal $\vec{x}$ when embedding a watermark associated to codevector $\vec{u}^{\star}$.} 
    		\label{fig:pushed}
	\end{figure}

	Ha gave a theoretical method to reach this value. He
	considered a signal $U \sim {\mathcal N}(0, P+\alpha^2 Q)$,
	know both from the embedder and the extractor. The capacity of
	the channel is then given by 	
	\begin{equation}		
		C = \max_{\alpha} \left\{ 
			I \left( U; Y \right) - 
			I \left( U; X \right)
		\right\} \textrm{,}
	\end{equation}
	where $Y \sim {\mathcal N}(0, Q+P)$ models the transmitted
	signal $\vec{y}$. Costa showed that the previous equation
	leads to the optimal value $\alpha = P/(P+N)$, and then to
	Eqn.~\ref{eq:costa}. The signal $U$ is obtained using a
	structured codebook of $2^{n\left( I(U; Y)- \epsilon \right)}$
	elements\footnote{With $\epsilon$ chosen to be very small as
	$n \rightarrow \infty$.}, designed to be a surjective function
	between the set of possible messages to embed ${\mathcal M}$
	and the codebook ${\mathcal U}$: each possible message
	$\vec{m}$ is associated to a sub-codebook ${\mathcal
	U}_{\vec{m}}$ composed of $2^{n I(U; X)}$ codewords. During
	the embedding process, the closest codeword $\vec{u}^{\star}
	\in {\mathcal U}_{\vec{m}}$ is chosen. The watermark signal is
	then given by 
	\begin{equation}
			\vec{w} = \vec{u}^{\star} - \alpha \vec{x}
			\textrm{.} \label{eq:alpha}
	\end{equation}
	Whereas classical watermarking techniques would have
	transmitted
	$\vec{x}+\sqrt{nP} \times \vec{u}^{\star} / \| \vec{u}^{\star} \|$,
	the $\alpha$ term forces the transmitted signal to go toward
	the codevector, as illustrated by Fig.~\ref{fig:pushed}. At
	the extraction process, the closest codeword
	$\widehat{\vec{u}} \in {\mathcal U}$ is computed. The decoded
	message is then $\widehat{\vec{m}}$ so that $\widehat{\vec{u}}
	\in {\mathcal U}_{\widehat{\vec{m}}}$.

	\subsection{Dirty paper codes from punctured ones}\label{sec:dirty}

	The original ICS is based on large random codebooks: the only
	way to decode $\vec{y}'$ is by an exhaustive search in ${\mathcal
	U}$. Some practical but suboptimal approaches, inspired by the
	ICS, have been proposed for i.i.d. Gaussian host signals,
	based on codebooks used for error correcting codes
	(ECC)~\cite{Ramkumar99,Eggers00b,ChenWornell01,Chou01}, where
	decoding process is designed to be much more simpler than
	an exhaustive search. 

\begin{figure}
  		\centerline{\epsfig{figure=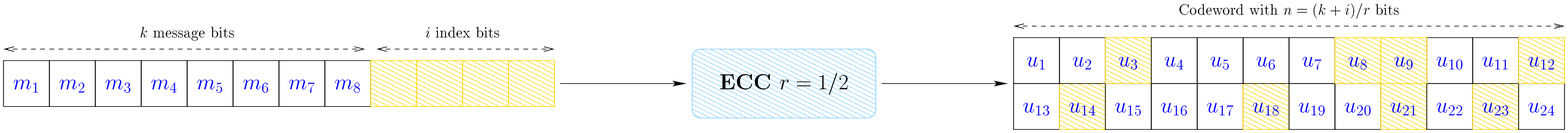,width=16cm}}
	    	\caption{Adding $i$ index bits to design a structured
  		codebook ($k=8$, $i=4$, $r=1/2$ leading to $n=24$).} 
    		\label{fig:index}
	\end{figure}

	Each possible $k$-long message is associated to $2^{n I(U; X)}$
	codewords. A simple way to design such a structured codebook
	would be to insert $i = n \times I(U; X)$ index bits in the
	message and to encode it. For an ECC with rate $r$, this
	leads to $n$-long codewords with $n=(k+i)/r$ (see
	Fig.~\ref{fig:index}). According to Costa, the value of $i$ is
	given by 
	\begin{equation}
		i = n \times I(U; X) = \frac{n}{2} \log_2 \left[
			1+ \frac{P Q}{\left( P+N \right)^2}
		\right] \textrm{,}
	\end{equation}
	which then depends on $Q$, {\it i.e.} the host signal. Thus the
	final codeword length $(k+i)/r$ may vary, while the host
	signal length $n$ is generally given and fixed (number of
	pixels for an image, sample size of a sound\ldots). The length
	of codewords must not depend on $i$, {\it i.e.} the global rate
	$k/n$ must be fixed. 

\begin{figure}
		\centering
			\subfigure[Construction of a pattern for the decoding
			of $\vec{x}$.\label{fig:pattern}]
			{
				\includegraphics[width=6cm]{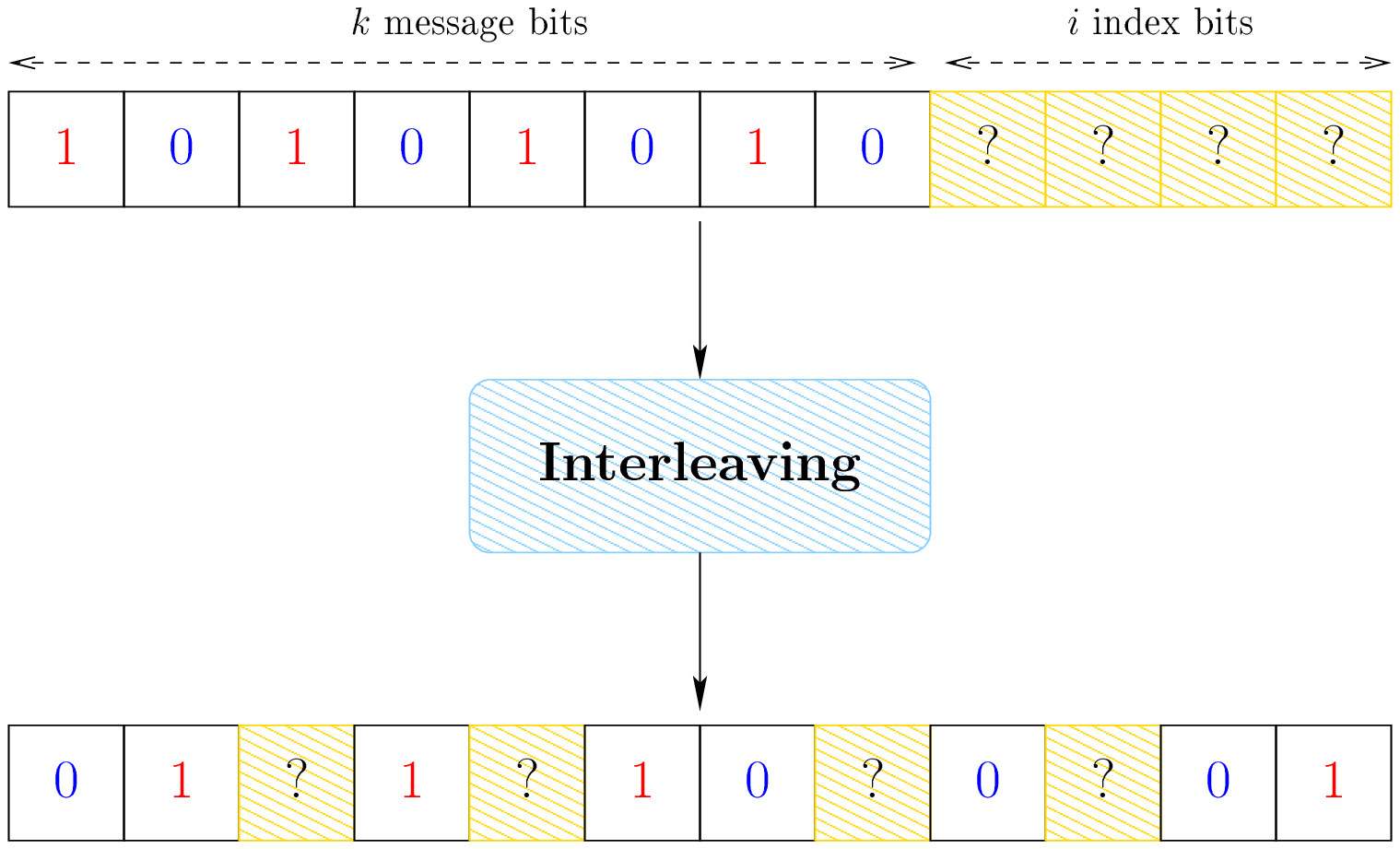}
			}
			\subfigure[Decoding the expanded host signal using a
	  		modified Viterbi algorithm.\label{fig:trellis}]
			{
				\includegraphics[width=12cm]{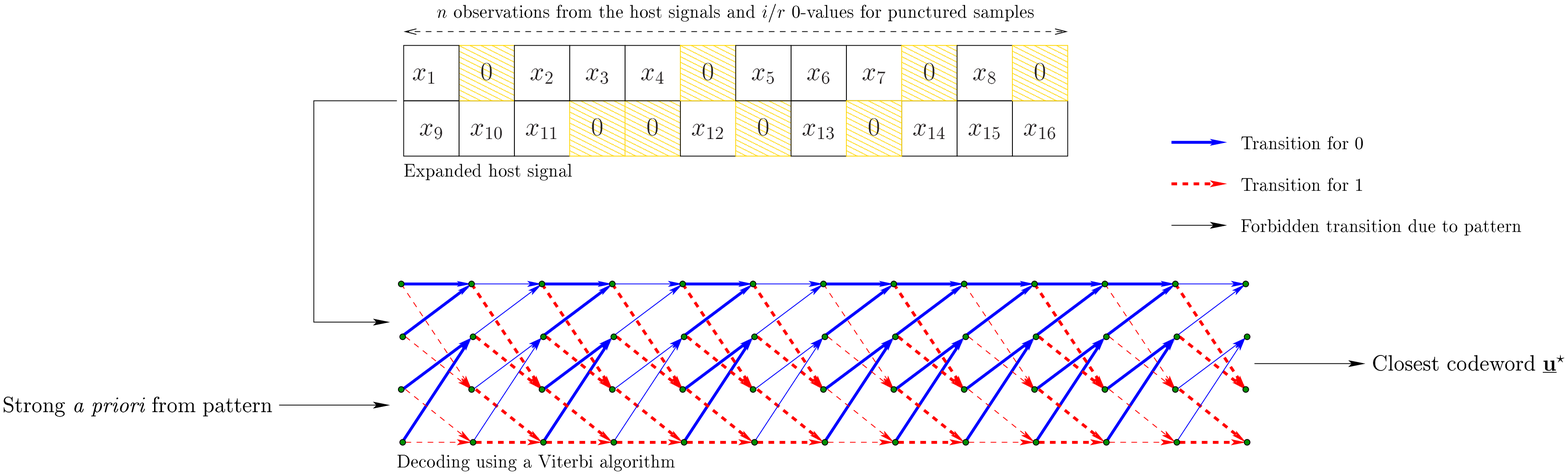}
			}
		\caption{The search for the closest codeword at the
		embedding stage ($k=8$,
		$i=4$, $n=16$ and $r=1/2$).}	
	\end{figure}

	We thus propose to use a simple codebook based on punctured
	convolutional codes and soft trellis decoding. Let us choose an
	error correcting code in order to get a rate $r = k/n$. 
	We then design an interleaved pattern composed of the $k$ bits
	from the message $\vec{m}$ to be embedded and of $i$ additional
	bits, as illustrated in Fig.~\ref{fig:pattern}. We then expand
	the host signal from $n$ to $(k+i)/r$ using neutral values for
	soft decoding ({\it i.e.} $0$). This expanded host signal is decoded
	with a modified soft Viterbi decoding algorithm, using 
	the previous $k$ bits pattern as a strong {\it a priori} in order to
	force some transitions in the convolutional trellis (see
	Fig.~\ref{fig:trellis}). The output fixes the $i$ index bits
	and gives a $(k+i)/r$-long codeword, which is punctured
	according to the previous expansion of the host signal, in
	order to remove $i/r$ bits and to finally get a $n$-long
	codeword. This leads to the closest codeword $\vec{u}^{\star}
	\in {\mathcal U}_{\vec{m}}$ to $\vec{x}$. Using
	BPSK\footnote{Binary Phase Shift Keying.}, all the obtained
	codewords are designed to have the same energy, {\it i.e.} $\|
	\vec{u} \| = \sqrt{n}$.  
	
	The watermark is finally chosen in order to get the
	maximum robustness~\cite{Miller00}: the codeword
	$\vec{u}^{\star}$ is associated to a hyper-cone of robustness,
	where $\vec{y}$ must lie into to be correctly decoded.
	Further, hyperboloids may be defined to represent set
	of points of given robustnesses ({\it e.g.} ${\mathcal H}_{N_1}$,
	${\mathcal H}_{N_2}$\dots on Fig.~\ref{fig:pushed}). The
	watermark $\vec{w}$ is defined in order to maximize
	robustness, that is 
	\begin{equation}
		\vec{w} = \arg \max_{\widetilde{\vec{w}}} \left\{
			\left[ 
				\frac{
					\left( 
						\vec{x} + \widetilde{\vec{w}}
					\right) \cdot \vec{u}^{\star}
				}{
					\| \vec{u}^{\star} \|
				}
			\right]^2 \left(
				1+ \tan^2 \theta
			\right) - \| \vec{x} + \widetilde{\vec{w}} \|^2
			\textrm{, with } \| \widetilde{\vec{w}} \| = \sqrt{nP}
		\right\} \textrm{,} \label{eq:mr}
	\end{equation}
	where $\theta$ is the angle of the hyper-cone, given by~\cite{Pateux03}
	\begin{equation}	
		\tan^{-2} \theta = 2^{	
			\frac{2\left( k+i \right)}{n}
		} -1\textrm{.}
	\end{equation}

	At the receiver, the signal $\vec{y}' = \vec{y}+\vec{z}$ is
	expanded from $n$ to $(k+i)/r$ elements insertion $0$
	elements, and decoded using the trellis to get $\vec{m}$.
	Thanks to soft decoding and to the fact that codewords have
	all same energy, this coding scheme is scale resistant, {\it
	i.e.} $\vec{y}'$ can be scaled ($\vec{y}' = \gamma \left[
	\vec{y} + \vec{z} \right]$ with $\gamma>0$) without loss of
	robustness. 

	\subsection{Game theory applied to spread spectrum\label{sec:game}} 

	Multimedia signals are not usually i.i.d. and Gaussian. So we
	consider a non i.i.d. host signal $\vec{x}$  modeled by a set
	of random variables $X^m = \left\{ X_1, X_2, \ldots, X_m
	\right\}$ with $X_i \sim {\mathcal N}(0, \sigma_{X_i}^2)$,
	{\it i.e.} signal is modeled as a mixture of Gaussians. We also
	consider a more general model for attack: SAWGN. The received
	signal can then be written as $y'_i = \gamma_i^{\textrm a}
	\times y_i + \delta_i$ where $\gamma_i^{\textrm a} \geq 0$ and $\delta_i$ is a Gaussian noise modeled by
	$Z_i \sim {\mathcal N}(0, \sigma_{Z_i}^2)$. To embed a $n$
	symbols length message in a $m$-long signal, wide spread
	spectrum uses a pseudo-random matrix $\vec{G} \in \left\{ -1; 1
	\right\}^{m \times n}$. This can be associated to a spread
	transform, {\it i.e.} the embedding process is made in a linear
	subspace, like for ST-DM~\cite{ChenWornell01} or
	ST-SCS~\cite{EggersSpie2001,EggBaumlGirod2002}. Our previous
	work~\cite{Guelvouit02,Guelvouit02b} demonstrated the interest
	of Wiener filtering at embedding\footnote{Since attacker would
	perform Wiener filtering to decrease $D_{xy'}$, Wiener
	filtering at embedding allows to decrease $D_{xy}$ without
	loss of performance.}: 
	\begin{eqnarray}
		y_i &=& \gamma_i^{\textrm w} \left[ 
			x_i + w_i 
		\right] = \gamma_i^{\textrm w} \left[ 
			x_i + \frac{\sigma_{W_i}}{
				\sqrt{nP}
			} \sum_{j=1}^n w_j^{\textrm {st}} 
			\times G_{i,j} 
		\right]	\label{eq:embed} \\
		\textrm{with } \gamma_i^{\textrm w} &=& \frac{
			\sigma_{X_i}^2
		}{
			\sigma_{X_i}^2 + \sigma_{W_i}^2
		} \textrm{.}
	\end{eqnarray}
	The watermark $\vec{w} = \left\{ w_1, w_2, \ldots , w_m \right\}$ is thus
	non i.i.d. and is modeled by $W^m$ with $W_i \sim {\mathcal
	N}(0, \sigma_{W_i}^2)$. The Wiener filtering and the scale
	attack can be grouped: $\gamma_i =
	\gamma_i^{\textrm a} \times \gamma_i^{\textrm w}$. 
	The inverse spread transform (used for extraction) is defined
	by a weighted linear correlation~\cite{Guelvouit02}: 
	\begin{eqnarray}
		x_j^{\textrm {st}} &=&
		\sum_{i=1}^m \beta_i \gamma_i \times x_i
		\times G_{i, j} \\ 
		\textrm{and } y'^{\textrm {st}}_j &=&
		\sum_{i=1}^m \beta_i \times y'_i
		\times G_{i, j} \textrm{,} 
		\label{eq:extract}
	\end{eqnarray}
	where $\beta_i$ is a weighting factor. As demonstrated
	previously~\cite{Pateux03} considering a SI scheme, the
	optimal value for $\beta_i$ can be expressed as 
	\begin{equation}
		\beta_i^{\star} \propto \frac{\gamma_i \times
		\sigma_{W_i}}{\sigma_{Z_i}^2} \textrm{.}
		\label{eq:beta}
	\end{equation}
	In this subspace, the embedding process from
	Eqn.~(\ref{eq:embed}) is written as 
	\begin{equation}
		\forall j \in \left\{ 1, 2, \ldots ,  n \right\} \textrm{, } y_j^{\textrm {st}} = x_j^{\textrm {st}} + w_j^{\textrm {st}} \textrm{,}
	\end{equation}
	where $\vec{x}^{\textrm {st}}$ is i.i.d. and
	Gaussian. We can then use Costa's approach
	described in Sec.~\ref{sec:costa}, and define from
	Eqns.~(\ref{eq:embed}) and~(\ref{eq:extract}) the different
	amounts of energy used as 
	\begin{eqnarray}
		Q &=& \sum_{i=1}^m \beta_i^2 \gamma_i^2 \times
		\sigma_{X_i}^2 \textrm{,} \label{eq:Q}\\
		N &=& \sum_{i=1}^m \beta_i^2 \times \sigma_{Z_i}^2
		\textrm{,} \label{eq:N}\\
		P &=& \frac{1}{n}  \left[ \sum_{i=1}^m
			\beta_i \gamma_i \times	\sigma_{W_i}
		\right]^2
		\textrm{.}
	\end{eqnarray}
	We remark that while $\sigma_{X_i}^2 \gg \sigma_{W_i}^2$ to
	ensure the invisibility of the watermark, the available
	watermark energy $P$ is concentrated in the subspace and
	can then become more important than the energy $Q$ of the host
	signal (when $m/n \gg 1$). It also shows that $P$ is shared by the symbols to
	be embedded: more symbols ({\it i.e.} larger $n$) means less watermark
	energy per symbol. 
	
	Given a maximum amount of embedding distortion, we must 
	optimize the embedding energy, {\it i.e.} $\sigma_{W_i}$. Define an
	embedding and an attack distortion functions: 
	\begin{eqnarray}
		D_{xy} &=& E \left[
			\varphi_i^2 \left(
				x_i - y_i
			\right)^2
		\right] = \frac{1}{m} \sum_{i=1}^m \varphi_i^2 \frac{
			\sigma_{X_i}^2 \sigma_{W_i}^2
		}{
			\sigma_{X_i}^2 + \sigma_{W_i}^2
		}\\
		\textrm{and } D_{xy'} &=& E \left[
			\varphi_i^2 \left(
				x_i - y'_i
			\right)^2
		\right] = \frac{1}{m} \sum_{i=1}^m \varphi_i^2 \left(
			\sigma_{X_i}^2 \left( 
				1- \gamma_i
			\right)^2 + \gamma_i^2 \sigma_{W_i}^2 +
			\sigma_{Z_i}^2
		\right) \textrm{,}
	\end{eqnarray}
	where $\varphi_i$ is a perceptual weighting factor. The performance of
	the inverse spread transform can be quantified by the
	signal-to-noise ratio $E_b/N_0$ defined as 
	\begin{equation}
		\frac{E_b}{N_0} = \frac{
			P
		}{
			N
		} = \frac{1}{n}
		\sum_{i=1}^m \frac{\gamma_i^2 \times
		\sigma_{W_i}^2}{\sigma_{Z_i}^2} \textrm{.}
	\end{equation}
	It should be noted that this value is not the signal-to-noise ratio
	obtained at the output of the extractor from
	Eqns.~(\ref{eq:extract}) and~(\ref{eq:beta}), given
	by~\cite{Pateux03} 
	\begin{equation}
		\textsf{snr} = \frac{
			P \left( P+Q+N \right)
		}{
			N \left( P+N \right)
		} \textrm{.}
	\end{equation}

	We now solve the optimization of $\sigma_{W_i}$ using a
	min-max game: given a maximal amount of distortion
	$D_{xy'}^{\max}$, the attacker wants to minimize $E_b/N_0$,
	while the embedder wants to maximize it, for a maximal amount
	of embedding distortion $D_{xy}^{\max}$. This is done by two
	Lagrangian optimizations~\cite{Guelvouit02}. First, for the
	attacker, we get the following functional: 
	\begin{equation}
		\forall i \in \left\{ 1, 2, \ldots ,  m \right\} \textrm{, }
		\left( 
			\gamma_i^{\star}, \sigma_{Z_i}^{\star}
		\right) = \arg \min_{\gamma_i, \sigma_{Z_i}} \left\{
			J_{\lambda, i} = \frac{
				\gamma_i^2 \times \sigma_{W_i}^2
			}{
				\sigma_{Z_i}^2
			} + \lambda \left[
				\varphi_i^2 \left(
					\sigma_{X_i}^2 \left( 
						1- \gamma_i
				\right)^2 + \gamma_i^2 \sigma_{W_i}^2 + \sigma_{\delta_i}^2
				\right)
			\right]	
		\right\} \textrm{,}
	\end{equation}
	where $\lambda$ is a Lagrangian multiplier used to respect the
	constraint on the attack distortion. This leads to the optimal
	values for $\gamma_i$ and $\sigma_{Z_i}$:
	\begin{eqnarray}
		\gamma_i^{\star} &=&  \frac{
			\sigma_{X_i}^2 - \frac{
				\sigma_{W_i}
			}{
				\varphi_i \sqrt{\lambda}
			}
		}{
			\sigma_{X_i}^2 +
			\sigma_{W_i}^2
		} \textrm{ if } \sigma_{W_i} \leq \sqrt{\lambda}
		\varphi_i \sigma_{X_i}^2 \nonumber \\
		&=& 0 \textrm{ otherwise,} \\
		\textrm{and } \left( \sigma_{Z_i}^{\star} \right)^2&=&
		\gamma_i^{\star} \left( 
			\gamma_i^{\textrm w} - \gamma_i^{\star}
		\right) \left( 
			\sigma_{X_i}^2 + \sigma_{W_i}^2
		\right) \textrm{.}
	\end{eqnarray}
	
	The second part of the game consists in optimizing the
	embedding parameters considering optimal attack, which is also
	done by a Lagrangian approach: 
	\begin{equation}
		\forall i \in \left\{ 1, 2, \ldots , m \right\} \textrm{, }
		\sigma_{W_i}^{\star} = \arg \max_{\sigma_{W_i}} \left\{
			J_{\chi, i} = J_{\lambda, i} - \chi \left[
				\varphi_i^2 \frac{
					\sigma_{X_i}^2 \sigma_{W_i}^2
				}{
					\sigma_{X_i}^2 + \sigma_{W_i}^2
				}
			\right]	
		\right\} \textrm{,}
	\end{equation}
	where $\chi$ is a Lagrangian multiplier used to respect the
	constraint on the embedding distortion. This leads to the
	final optimal embedding parameters 
	\begin{equation}
		\sigma_{W_i}^{\star} = \frac{
			\varphi_i^2 \left( \lambda - \chi \right) \sigma_{X_i}^2 -1 +
			\sqrt{
				\left(
					\varphi_i^2 \left(
						\lambda-\chi 
					\right) \sigma_{X_i}^2 -1
				\right)^2 +4 \lambda \varphi_i^2 \sigma_{X_i}^2
			}
		}{
			2 \sqrt{\lambda} \varphi_i
		} \textrm{,}
	\end{equation}
	and also to a particular expression for the optimal correlation
	factor for inverse spread transform: $\beta_i^{\star} \propto
	\varphi_i$, when considering optimal attacks. 

\section{INTERFERENCE CANCELLATION}\label{sec:ic}

\begin{figure}
  		\centerline{\epsfig{figure=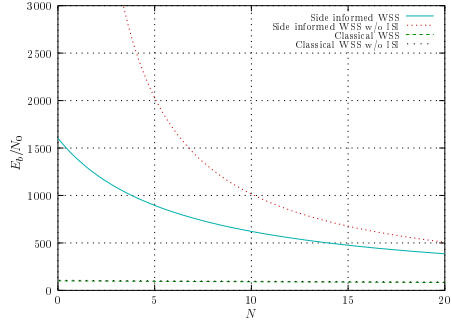,width=6cm}}
	    	\caption{Signal-to-noise ratio against AWGN attack,
  		for the classical image {\it Lena}
  		($n=162$, $m=512 \times 512$ and $E\left[
  		\sigma_{W_i} \right] = 2.5$ $\forall i \in \{1, 2,
  		\ldots , m \} $).} 
    		\label{fig:noise}
	\end{figure}

	Without side informed watermarking, the signal-to-noise ratio
	we get is given by $E_b/N_0 = P/(Q+N)$. In practice, this
	value is correct only if the carriers $\vec{G} = \left\{
	\vec{G}_0, \vec{G}_1, \ldots , \vec{G}_n \right\}$ of the
	spread transform are truly orthogonal, which is not the case
	with pseudo-random carriers. Thus the signal-to-noise ration
	is given by 
	\begin{eqnarray}
		\frac{E_b}{N_0} &=& \frac{P}{Q+N+I} \\
		\textrm{where } I &=& \sum_{i=1}^m
		\beta_i^2 \gamma_i^2 \times 
		\frac{\sigma_{W_i}^2}{
			nP
		} \left(
			n-1
		\right) \textrm{.}
	\end{eqnarray}
	The value $I$ is known as the inter-symbols interference. In
	non-informed watermarking techniques (where the host signal
	influences the performance of the scheme), this interference is
	negligible because $Q \gg I$. But in informed
	watermarking, for a low level of attack, it represents a great
	amount of noise that limits the robustness and/or the capacity
	of the scheme. Fig.~\ref{fig:noise} illustrates the gap
	between WSS watermarking with pseudo-random carriers and
	theoretical WSS watermarking (with truly orthogonal carriers).
	We will see in the remaining part of this section three
	methods to cancel this interference. 

	\subsection{Insuring orthogonality of the carriers}

	To avoid interference, a trick is to embed only one symbol per
	host element~\cite{ChenWornell01}, {\it i.e.} $\forall i \in
	\left\{ 1, 2, \ldots m \right\}$, there is only one element in $\left\{ G_{i, 1}, G_{i, 2},
	\ldots , G_{i, n} \right\}$ which is not set to $0$. In this
	case, $I=0$. However this technique limits the spreading of the
	bits, especially for important values of $n$, case where
	interference cancellation is very interesting (low level of
	attack noise). 

\begin{figure}
		\centering
			\subfigure[Original image.]
			{
				\includegraphics[width=5cm]{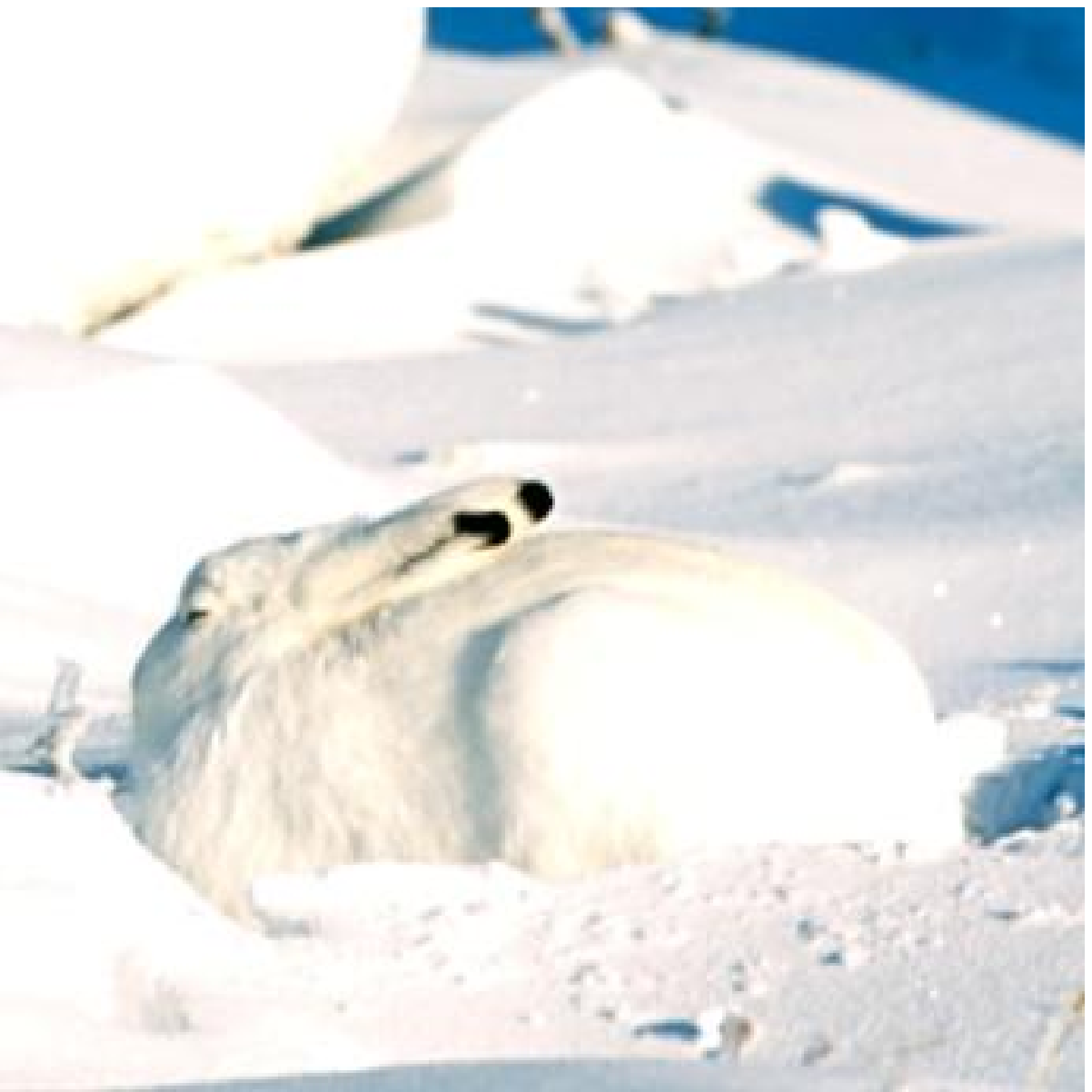}
			}
			\subfigure[Tree-levels DWT transform of {\it
			Artic hare}.]
			{
				\includegraphics[width=5cm]{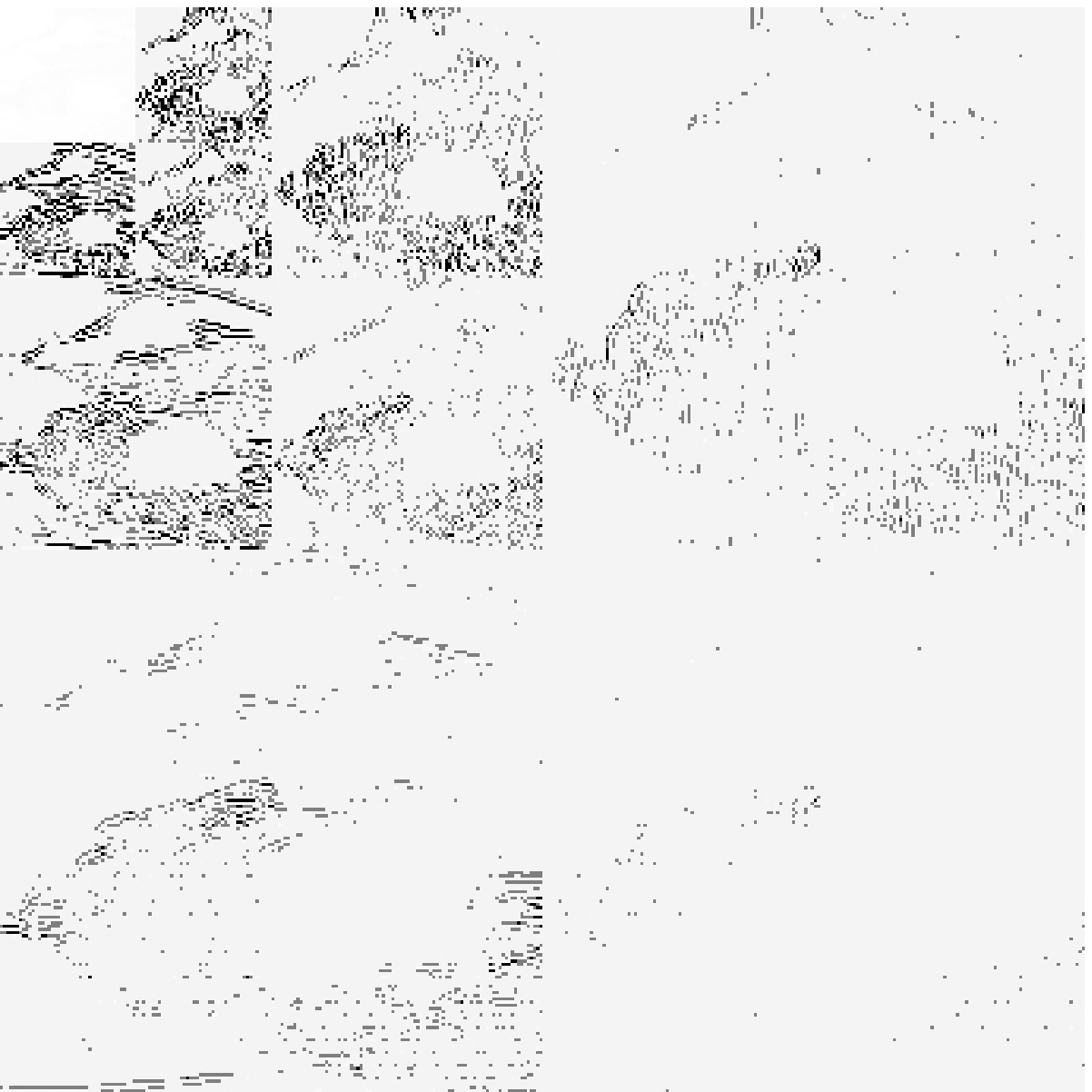}
			}
		\caption{{\it Artic hare}, a difficult image to watermark: the number of
  		interesting elements is limited (copyright photos
  		courtesy of Robert E.~Barber, Barber Nature
  		Photography).}	
		\label{fig:artic}
	\end{figure}

	Moreover, in the case of smooth signals (see
	Fig.~\ref{fig:artic} for an example), the number of
	well suited host elements (important value of $\gamma_i
	\sigma_{W_i} / \sigma_{X_i}$) is limited. The energy
	of the watermark is mainly located on high
	energy coefficients. Since  this number of coefficients is 
	small, symbols to be hidden may not be equally spread
	over the host signal ({\it i.e.} a symbol may be spread on non
	significant coefficients whereas an other one will be spread
	on significant ones). It results in the linear subspace in non
	i.i.d. signals. Performances are not guaranteed and parallel
	channels should rather be considered. 

	\subsection{Cancellation at the decoder}

	If pseudo-random carriers were used at embedding, the received
	signal in the spread transform subspace can be written as 
	\begin{equation}
		\vec{y}'^{\textrm {st}} = \gamma \left[ 
			\vec{x}^{\textrm {st}} + \vec{w}^{\textrm
			{st}} + \textsf{isi}(
				\vec{w}^{\textrm {st}}
			)
		\right] + \vec{\delta}^{\textrm {st}} \textrm{,}
	\end{equation}
	where $\textsf{isi}(\vec{w}^{\textrm {st}})$ is
	the inter-symbols interference. For $Q \ll P$ (very common case for
	payloads such as $n<1000$), we can write $\vec{x}^{\textrm
	{st}} + \vec{w}^{\star} \simeq \vec{w}^{\textrm {st}} \simeq
	\sqrt{P} \times \vec{u}^{\star}$ and then 	
	\begin{equation}
		\vec{y}'^{\textrm {st}} \simeq \gamma \left[ 
			\vec{u}^{\star} + \textsf{isi}(
				\sqrt{P} \times \vec{u}^{\star}
			)
		\right] + \vec{\delta}^{\textrm {st}} \textrm{.}
	\end{equation}
	Thus we can estimate $\textsf{isi}( \sqrt{P} \times \vec{u}^{\star}
	)$ in order to cancel ISI. Receiver first estimates
	$\vec{u}^{\star}$. Corresponding interference is then canceled. The new
	$\vec{y}'^{\textrm {st}}$ is obtained and used to compute
	$\widetilde{\vec{u}}^{\star}$. This process iterates until
	$\vec{u}^{\star} = \widetilde{\vec{u}}^{\star}$ (see
	Alg.~\ref{alg:post}). To be efficient, receiver must know (or
	estimate) embedding energy $\sigma_{W_i}$. Moreover, optimal
	scaling factor $\gamma_i^{\star}$ must also be estimated. This
	may be done by an additional reference signal, leading to a
	lower capacity for message bits. We will then search for
	another solution consisting in canceling ISI at embedding.
	\begin{algorithm}
		\caption{Considering $\vec{w}^{\textrm {st}} \simeq
		\sqrt{P} \times \vec{u}^{\star}$, search for the closest codeword
		$\vec{u}^{\star}$ from $\vec{y}'^{\textrm {st}}$
		with ISI canceled}
		\label{alg:post}
	\begin{algorithmic}	
		\FOR{$j=1$ to $n$}
			\STATE $y'^{\textrm {st}}_j$ $\leftarrow$ 
			$\displaystyle \sum_{i=1}^m \beta_i \times y'_i \times G_{i,
			j}$
		\ENDFOR
		\STATE

		\STATE $\widetilde{\vec{u}}^{\star}$ $\leftarrow$ closest codeword
		to $\vec{y}'^{\textrm {st}}$

		\STATE 
		\REPEAT
			\STATE $\vec{u}^{\star}$
			$\leftarrow$ $\widetilde{\vec{u}}^{\star}$
			\STATE

			\FOR{$j=1$ to $n$}
				\STATE $y'^{\textrm {st}}_j$ $\leftarrow$ $0$
				\FOR{$i=1$ to $m$}
					\STATE $I_{i, j}$ $\leftarrow$
					$\displaystyle \gamma_i \frac{
						\sigma_{W_i}
					}{
						\sqrt{nP}
					} \sum_{k=1 \textrm{, }k\neq
					j}^n \left(
					\widetilde{u}_k^{\star} \times G_{i,
					k} \right)$
		
					\STATE $y'^{\textrm {st}}_j$ $\leftarrow$
					$\displaystyle y'^{\textrm {st}}_j +
					\beta_i \left(		
						y'_i - I_{i, j}
					\right) G_{i, j}$
				\ENDFOR
			\ENDFOR

			\STATE
			\STATE $\widetilde{\vec{u}}^{\star}$ $\leftarrow$ closest
			codeword to $\vec{y}'^{\textrm {st}}$
				
		\UNTIL{$\vec{u}^{\star} = \widetilde{\vec{u}}^{\star}$}
	\end{algorithmic}
	\end{algorithm}

	\subsection{Interference as side information}

	As seen in Sec.~\ref{sec:costa}, the use of the side
	information available during the embedding process leads to
	great improvements, and if no attack is applied during the
	transmission, the capacity of the channel is infinite. But the
	spread transform we use to embed the watermark introduces a
	noise that limits capacity due to ISI. We propose to
	consider ISI as a kind of side information. 

	Even if this interference is introduced by the embedder, it is
	not perfectly known before the embedding. So, it can not be
	directly considered as side information. The problem is that the
	interference depends on the watermark signal, which depends on
	the interference. We use an iterative
	algorithm to converge to a watermark signal that takes into
	account its own interference. This algorithm is described by
	Alg.~\ref{alg:ic}. We first compute $\vec{w}^{\textrm {st}}$,
	as explained in Sec.~\ref{sec:state}. The interference it
	produces is computed, and introduced as side
	information. In a second step, this new side information
	$\widetilde{\vec{x}}^{\textrm {st}}$ is used to compute an
	updated watermark signal. The previous steps are iterated
	until $\vec{w}^{\textrm {st}}$ converges (we 
	observe convergence is attained to after typically less than 3
	iterations). At the end of the loop, the watermark signal
	takes into account the host signal and the symbol
	interference, and is added using Eqn.~(\ref{eq:embed}). 

	\begin{algorithm}
		\caption{Calculate $\vec{w}^{\textrm{st}}$ considering
		ISI as side information}
		\label{alg:ic}
	\begin{algorithmic}	
		\FOR{$j=1$ to $n$}
			\STATE $x_j^{\textrm {st}}$ $\leftarrow$ 
			$\displaystyle \sum_{i=1}^m \beta_i
			\gamma_i^{\star} \times x_i \times G_{i, j}$
		\ENDFOR
		\STATE

		\STATE $\vec{u}^{\star}$ $\leftarrow$ closest codeword
		to $\vec{x}^{\textrm {st}}$
		
		\STATE $\widetilde{\vec{w}}^{\textrm {st}}$
			$\leftarrow$ $\displaystyle \arg \max_{\vec{w}^{\textrm {st}}} \left\{
				\left[ 
					\frac{
						\left( 
							\vec{x} + \vec{w}^{\textrm {st}}
						\right) \cdot \vec{u}^{\star}
					}{
						\| \vec{u}^{\star} \|
					}
				\right]^2 \left(
					1+ \tan^2 \theta
				\right) - \| \vec{x} + \vec{w}^{\textrm {st}} \|^2
				\textrm{, with } \| \vec{w}^{\textrm {st}} \| = \sqrt{nP}
			\right\}$

		\STATE 
		\REPEAT
			\STATE $\vec{w}^{\textrm {st}}$
			$\leftarrow$ $\widetilde{\vec{w}}^{\textrm {st}}$
			\STATE

			\FOR{$j=1$ to $n$}
				\STATE $x_j^{\textrm {st}}$ $\leftarrow$ $0$
				\FOR{$i=1$ to $m$}
					\STATE $w_i$ $\leftarrow$
					$\displaystyle
					\frac{\sigma_{W_i}}{\sqrt{nP}}
					\sum_{k=1}^n w_k^{\textrm
					{st}} \times G_{i, k}$

					\STATE $I_{i, j}$ $\leftarrow$
					$\displaystyle w_i -
					\widetilde{w}_j^{\textrm {st}}
					\times \frac{
						\sigma_{W_i} 
					}{
						\sqrt{nP}
					} \times G_{i, j}$

					\STATE $x_j^{\textrm
					{st}}$ $\leftarrow$
					$\displaystyle x_j^{\textrm
					{st}} + \beta_i \gamma_i^{\star}
					\left(
						x_i + I_{i, j}
					\right) G_{i, j}$
				\ENDFOR			
			\ENDFOR

			\STATE
			\STATE $\vec{u}^{\star}$ $\leftarrow$ closest
			codeword to $\vec{x}^{\textrm {st}}$
	
			\STATE $\widetilde{\vec{w}}^{\textrm {st}}$
				$\leftarrow$ $\displaystyle \arg \max_{\vec{w}^{\textrm {st}}} \left\{
				\left[ 
					\frac{
						\left( 
							\vec{x} + \vec{w}^{\textrm {st}}
						\right) \cdot \vec{u}^{\star}
					}{
						\| \vec{u}^{\star} \|
					}
				\right]^2 \left(
					1+ \tan^2 \theta
				\right) - \| \vec{x} + \vec{w}^{\textrm {st}} \|^2
				\textrm{, with } \| \vec{w}^{\textrm {st}} \| = \sqrt{nP}
			\right\}$
			
		\UNTIL{$|\widetilde{\vec{w}}^{\textrm
		{st}}-\vec{w}^{\textrm {st}}| \leq \epsilon$}
	\end{algorithmic}
	\end{algorithm}

\section{EXPERIMENTAL RESULTS}\label{sec:results}

	The previous studies have been applied to image watermarking.
	A 3-levels wavelet transform of a gray-scale image generates
	the host signal $\vec{x}$ ($m$ is equal to the
	number of pixels of the host image). We embed $k=64$ bits
	using a structured codebook, as described in
	Sec.~\ref{sec:dirty}, with a rate equal to $1/2$. This leads to
	$n=132$ with some padding bits. We consider a psycho-visual
	factor inspired from Watson's~\cite{Watson93} model, defined by 
	\begin{equation}
		\varphi_i = \frac{
			\rho
		}{
			\sqrt{
				\overline{\sigma_{X_i}} +1
			}
		} \textrm{,}
	\end{equation}
	where $\rho$ is set to get $E\left[ \varphi_i \right]=1$ and
	$\overline{\sigma_{X_i}}$ is a normalized activity measure
	(based on the variance of $X_i$). We finally tune $\lambda$
	and $\chi$ to obtain an embedding distortion equal to $7.0$
	({\it i.e.} $\textsf{wpsnr}(\vec{x}, \vec{y}) = 39.7$~dB). Two
	attacks are tested: Gaussian noise and JPEG lossy compression.

\begin{figure}
		\centering
			\subfigure[Performance against AWGN attack.\label{fig:lena:noise}]
			{
				\includegraphics[width=6cm]{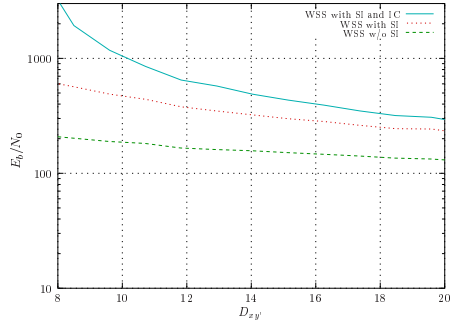}
			}
			\subfigure[Performance against JPEG lossy
			compression ($D_{xy'}=7$ for 100~\% JPEG quality, and
			$D_{xy'}\simeq 20$ for 15~\% JPEG quality).\label{fig:lena:jpeg}]
			{
				\includegraphics[width=6cm]{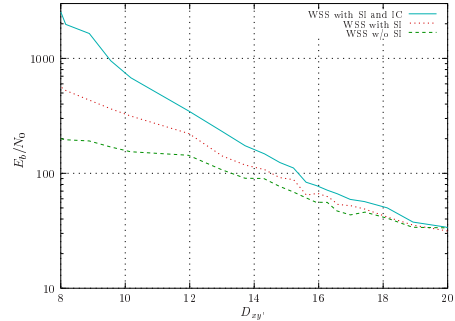}
			}
		\caption{Signal-to-noise ratio against attacks for {\it Lena} ($512 \times 512$ gray-scale image, 3-levels DWT, $n=132$ and $D_{xy}=7$).}	
		\label{fig:lena}
	\end{figure}

\begin{figure}
		\centering
			\subfigure[Performance against AWGN attack.\label{fig:paper:noise}]
			{
				\includegraphics[width=6cm]{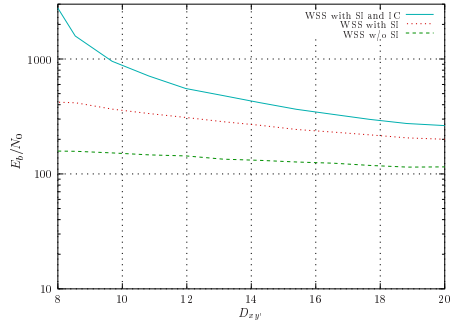}
			}
			\subfigure[Performance against JPEG lossy
			compression ($D_{xy'}=7$ for 100~\% JPEG quality, and
			$D_{xy'}\simeq 20$ for 15~\% JPEG quality).\label{fig:paper:jpeg}]
			{
				\includegraphics[width=6cm]{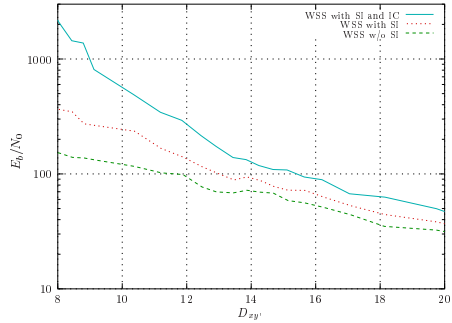}
			}
		\caption{Signal-to-noise ratio against attacks for
		{\it Paper machine} ($512 \times 512$ gray-scale image, 3-levels DWT, $n=132$ and $D_{xy}=7$).}	
		\label{fig:paper}
	\end{figure}

	For each attack level (energy of the added noise for AWGN
	attack and quality factor for JPEG compression), the resulting
	distortion $D_{xy'}$ is computed and the watermark is
	extracted to get the signal-to-noise ratio $E_b/N_0$.
	Figs.~\ref{fig:lena} and~\ref{fig:paper}\footnote{Both
	used images are available from F.~Petitcolas' web site:
	\url{<http://www.cl.cam.ac.uk/~fapp2/watermarking/image_database>}.} 
	confirm the interest of interference cancellation, already
	shown by the theoretical Fig.~\ref{fig:noise}. 

\section{CONCLUSION} \label{sec:conclusion}

	We studied in this paper a practical implementation of a
	watermarking scheme exploiting side information.
	We propose a method scheme based on a simple
	structured codebook using a soft Viterbi decoder. A
	spread transform gets i.i.d. signals from non i.i.d. ones.
	Embedding in the linear subspace defined by the spread
	transform generates inter-symbols interference. An iterative
	algorithm estimates this interference and includes it into the
	side information. We finally applied this scheme to image
	watermarking: this leads to important improvements in term of
	capacity and/or robustness. 

\bibliography{spie}
\bibliographystyle{spiebib} 

\end{document}